\documentclass{article}

\usepackage{arxiv}

\usepackage[utf8]{inputenc}
\usepackage{hyperref}
\usepackage[T1]{fontenc}
\usepackage{url}
\usepackage{booktabs}
\usepackage{amsfonts}
\usepackage{nicefrac}
\usepackage{microtype}
\usepackage{graphicx}
\usepackage{siunitx}
\usepackage{amsmath}
\usepackage{textcomp}

\title{Kinetic and chemorheological modeling of thermosetting polyurethanes obtained from an epoxidized soybean oil polyol crosslinked with glycerin}

\author{ \href{https://orcid.org/0000-0002-7342-1784}{\includegraphics[scale=0.06]{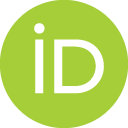}\hspace{1mm}Leonel M.~Chiacchiarelli}\thanks{ lmchiacchiarelli@yahoo.com.ar, TEL/FAX: +541143043020, Av. Gral Las Heras 2214, AAR1127, Buenos Aires, Argentina} \\
	Instituto de Tecnología de Polímeros y Nanotecnología (ITPN)\\
	CONICET-UBA,\\
	Buenos Aires, Argentina. \\
	Instituto Tecnológico de Buenos Aires,\\
	Departamento de Ingeniería Mecánica,\\
	Av. E. Madero 399, Buenos Aires, Argentina.\\
	\And
	\href{https://orcid.org/0000-0003-1221-2994}{\includegraphics[scale=0.06]{orcid.png}\hspace{1mm}Franco ~Armanasco} \\
	Instituto de Tecnología de Polímeros y Nanotecnología (ITPN)\\
	CONICET-UBA,\\
	Buenos Aires, Argentina. \\
	Instituto Tecnológico de Buenos Aires,\\
    Departamento de Ingeniería Mecánica,\\
	Av. E. Madero 399, Buenos Aires, Argentina.\\
	\And
	Sebastián ~D'hers\\
	Instituto Tecnológico de Buenos Aires,\\
	Departamento de Ingeniería Mecánica,\\
	Av. E. Madero 399, Buenos Aires, Argentina.\\
}

\hypersetup{
pdftitle={Kinetic and chemorheological modeling of thermosetting polyurethanes obtained from an epoxidized soybean oil polyol crosslinked with glycerin},
pdfsubject={polyurethane},
pdfauthor={Leonel M. ~Chiacchiarelli},
pdfkeywords={thermosetting polyurethane, epoxidized soybean oil, glycerin, chemorheology, cure kinetics},
}

\begin{document}
\maketitle

\begin{abstract}
	  Thermosetting polyurethanes were obtained using an aromatic isocyanate and a hydrophobic polyol formulation obtained from epoxidized soybean oil (ESO) crosslinked with glycerin. A systematic DSC analysis of the effect of catalyst type, crosslinker concentration, isocyanate index and ESO crystallization on the cure kinetics was conducted. The combination of a stannic catalyst at 0.2 wt.\% and glycerin at 20 wt.\% produced a cure kinetics governed by autocatalytic heat flow where vitrification played a key role in the formation of chemical bonds.  The evolution of Tg as a function of conversion, which followed Di-Benedetto’s predictions, supported the hypothesis that vitrification was a preponderant phenomenon during cure. The ESO crystallization had a melting endotherm centered at 35°C, affecting substantially the availability of reactive hydroxide groups. Dynamic Mechanical Analysis (DMA) of a post-cured sample revealed a Tg centered at 220°C, whereas quasi-static flexural mechanical tests revealed a flexural modulus of 2.14 GPa and a flexural strength of 99.4 MPa. Rheological experiments using a rheometer at isothermal conditions supported the hypothesis that vitrification played a key role in the evolution of apparent viscosity. A master model using Kim-Macosko equations was obtained for the proposed formulation.
\end{abstract}


\section{Introduction}
The development of new applications based on innovative thermoset polymers is strongly correlated to the understanding of its chemorheological behavior \cite{fredi2019novel,ageyeva2019review,van2007reactive}. Obtaining a chemorheological master model \cite{chiacchiarelli2012cure,kenny1990model,lelli2009,maffezzoli1994correlation} is a key prerequisite for the successful application of polymers in advanced applications. For example, polymer composite prepregs \cite{mazumdar2001composites} represent a material of choice for advanced applications in aeronautics as well as aerospace applications \cite{mazumdar2001composites,mallick2007fiber}. Taking into account that the control and evolution of chemical conversion is a key issue for the development of prepregs, it is crucial to use chemorheological master models to that purpose.  This, in turn, helps to avoid costly experiments whole sole purpose is to obtain an empirical solution to the problem. Even though several studies have been conducted to understand the chemorheological behavior of polyurethane thermosets \cite{chiacchiarelli2012cure,chiacchiarelli2013kinetic,diaz2011cure,milanese2011cure,papadopoulos2009thermal,rodrigues2005dsc,saha2011study}, the authors have  not found any focusing on thermosets synthesized from soybean oil.\\
	\\The search of sustainable and environmentally friendly solutions within the polyurethane industry has fueled scientific and industrial activities which emphasize on the replacement on fossil-fuel precursors with renewable ones \cite{chiacchiarelli2019sustainable}. One of the most relevant initiatives in this regard is the development of polyols from vegetable oils \cite{desroches2012vegetable,fridrihsone2020life,miao2014vegetable,petrovic2008polyurethanes,xia2010vegetable,petrovic2013biological}. Within the Americas region, soybean-oil based polyols are currently being developed both in industry and in scientific laboratories with the general objective to partially or fully replace petroleum-based polyols. Thermosetting polymers are ubiquitous for the development of polymer composites, however, a highly crosslinked network is a requirement to achieve that goal \cite{ionescu2005chemistry}. In polyurethanes, this means that highly functional polyols as well as isocyanates are mandatory. As we have already pointed out in a previous publication \cite{herran2019highly}, the synthesis of high hydroxide number (OH\#) biobased polyols combined with a suitable functionality (>3.0) and low apparent viscosity (< 2000 cp) represents a challenge within this  field. However, these polyols have a key property which renders it suitable for thermosetting applications, that is, hydrophobicity. Epoxidation of soybean-oil is a frequent intermediate synthesis route to obtain polyols with low OH\# \cite{petrovic2002epoxidation,campanella2006,campanella2007}. A slight change of temperature \cite{petrovic2002epoxidation,campanella2006,campanella2007,campanella2008,santacesaria2020soybean,aguilera2018epoxidation} within the synthesis protocol can lead to reduced epoxidation efficiencies, giving rise to a polyol with a low OH\#, good functionality and relatively low apparent viscosity \cite{herran2019highly}. In addition, the presence of oxirane moieties as well as low polar fatty ester chains renders the polyol a high hydrophobic character. It is known that polyurethane formulations are prone to the formation of gases during cure \cite{szycher1999handbook}. Interestingly, a hydrophobic polyol can be extremely useful to circumvent this issue, allowing the development of a thermosetting which can potentially be applied to polymer composites. Introducing this requirement in the analysis and taking into account that these polyols have an OH\# within the order of 150 mg KOH.mg$^{-1}$ and functionalities of around 2.9, then, it is essential to have a crosslinker to develop a suitable formulation. We propose glycerin as a suitable candidate to achieve this goal. Being a byproduct of the biodiesel industry makes this biobased crosslinker a key component to improve the thermomechanical properties of the final polymer. Even though glycerin has been used previously as a crosslinker as well as a starter for polyol synthesis \cite{anbinder2020structural,meiorin2020comparative,corcione2009glass,czachor2021hydrophobic}, the authors have not found any study using the formulation proposed in this study .\\
	\\In this work, a complete chemorheological analysis of a thermosetting polyurethane (TsPU) which is suitable for its application in polymer composites has been developed using differential scanning calorimetry (DSC) as well as rotational rheometry. The effect of catalyst type and concentration, crosslinker concentration, isocyanate index (NCO$_{index}$) as well as the polyol crystallization has been systematically incorporated in the model. Thermomechanical studies of uncured and post-cured samples using dynamical mechanical thermal analysis (DMA) and quasi-static flexural bending were performed. Finally, a general chemorheological master model is developed for the proposed formulation.\\ 
	  
\section{Materials and Methods}
\label{sec:headings}

\subsection{Materials and sample preparation procedures}
	The epoxidized soybean oil polyol (ESO) was obtained using an epoxidation protocol based on formic acid (Anedra, A.C.S. grade) as the carrier and hydrogen peroxide (Stanton, 30\% vol.)  as the oxidizer, following the procedure of previous works \cite{petrovic2002epoxidation,campanella2006,campanella2007,campanella2008}. The reactions were performed at isothermal conditions (50\textdegree\ C) using a RBD grade soybean oil provided by Molinos Rio de La Plata and by mixing initially the soybean oil with formic acid and subsequently adding the oxidizer at a rate of 2 ml.min$^{-1}$. The organic phase was extracted using ethyl acetate (Biopack, A.C.S. grade) and neutralization with a saturated solution of NaCl. Finally, the resultant polyol, from now on denominated SOYPOL, was degassed using a vacuum mixer for 2 hours. The OH\# of the resultant polyol (137 mgKOH.mg$^{-1}$) was obtained using the guidelines of the test method A provided by the ASTM D4274.\\
	\\A polymeric methylene diphenyl diisocyanate (pMDI), denominated commercially as Suprasec 5005 (Huntsman), with an NCO$_{number}$ of 31.0 and a functionality of 2.7 was used to prepare all the thermosetting samples. Before use, the pMDI was degassed using a dispermat vacuum mixer at 30 mbar and stirring at 5300 rpm. Glycerin (Anedra, USP) with a functionality of 3.0 and OH\# of 1800 mgKOH.mg$^{-1}$ was used after degassing in a vacuum mixer. Dibutyltin dilaurate (95\%, Sigma Aldrich) and dimethylcyclohexylamine (99\%, Huntsman) were used as catalysts.\\
	\\The polyurethane formulations used in this work are reported in Table 1. A total of eight formulations were tested to analyze the effect of catalyst type, isocyanate index (NCO$_{index}$) and crosslinking. The general preparation procedure consisted first with the degassing of the components using the protocols described in the last paragraph. Then, samples of approximately 80 g were prepared by mixing all the components listed in Table 1 and by adding the pMDI as the final one. Finally, the sample was mixed at a speed of 7800 rpm at 30 mbar. It is important to highlight that, only for section 4, the final addition of pMDI was performed using manual mixing. As already noticed in previous studies \cite{castro1982studies,macosko1989rim}, the mixing intensity has a fundamental role in the reactivity of polyurethane formulations. DSC analysis performed with hand mixed samples of the TSPU$_{7}$ revealed the presence of both low and high temperature transitions, indicating that the mixing efficiency lead to a probable reaction between the isocyanate radicals.\\

\subsection{Sample characterization techniques}
	Heat flow analysis was performed using a Shimadzu DSC-60 differential scanning calorimeter (DSC) equipped with Aluminum hermetic pans. The dynamic thermal cycles started at 25\textdegree\ C going up to 200\textdegree\ C at a scan rate of 2\textdegree\ C.min$^{-1}$. Sample mass was typically within the order of 10 mg.\\
	\\FTIR spectra were obtained using a Shimadzu IRAffinity-1 using attenuated total reflection (ATR) and absorption methods. The ATR method was used for liquid samples (uncured precursors). The ATR consisted of a Miracle single reflection device equipped with a ZnSe prism. Each spectrum was obtained by recording 50 scans in the range 4000 cm$^{-1}$ to 600 cm$^{-1}$ with a resolution of 4 cm$^{-1}$ at ambient temperature. The phenyl absorption band (centered at 1595 cm$^{-1}$) was used to height normalize each spectra.\\
	\\Dynamic Mechanical Thermal Analysis (DMA) was carried out using a Perkin \& Elmer DMA 8000 equipment using the single cantilever bending fixture mode. The oscillation frequency was fixed at 1.0 Hz and the amplitude to 0.004 mm, previously selected from a strain scan at ambient temperature. At least three samples were used to corroborate the reproducibility of the results. The sample dimensions were in the order of 10 mm in length, 8.8 mm in width and 1.8 mm in thickness.\\
	\\Rheological measurements were performed using a Brookfield Viscometer (Myr VR 3000) with a LV-3 stainless steel spindle. The measurements were performed at isothermal conditions with samples of approximately 150 gr. using a stainless steel container submersed in an oil bath thermally controlled with a heating plate (Velp Scientific).\\
	\\Quasi-static flexural mechanical tests were performed with an Instron 5985 following the guidelines of the standard ASTM D790. For each mechanical test, five samples were tested. The span to depth ratio was set to 16\: 1 and the speed of the flexural deformation was 0.1 mm.min$^{-1}$. These samples were post-cured in a Horizontal Forced Air Drying Oven (Milab 101-2AB) before flexural tests from ambient temperature to 70\textdegree\ C and to 110\textdegree\ C at 10\textdegree\ C.min-1 with heating steps of 20\textdegree\ C and 60 min. dwell time. Both post-cure cycles were concluded with an outdoor cooling.\\

\section{Results and discussion}
\subsection{The role of catalyst type and concentration on the cure of the TsPU}
	A key issue for the development of a TsPU formulation is the proper selection of catalyst type and concentration. If the main objective is to develop a thermosetting polyurethane, then, it is desirable that the catalyst should selectively promote the formation of urethane or other covalent bonds (gelling chemical reactions) while simultaneously avoiding the formation of CO$_{2}$(g), which is caused mainly by the reaction of isocyanate with water (blowing reaction). Organotin compounds, such as stannous octoate and dibutyltin dilaurate (DBTDL), are currently widely employed in industry to this effect \cite{Randall2002ThePB}. However, the main drawback of such catalysts are associated to increasing the toxicity of the resultant polyurethane formulation, particularly for biomedical applications. Organocatalysts \cite{sardon2015synthesis} have a great potential to replace organotin compounds, however, their availability in industry is still limited to niche applications. Tertiary amines, such as DMCHA, have also been widely employed to alleviate this effect, however, its selectivity towards blowing reactions is much higher than gelling reactions. This represents a serious drawback because TsPUs need to have very low porosity levels to achieve improved thermomechanical properties. When it comes to the concentration of the catalyst, the minimum value is usually associated to the minimum amount which is capable to bring a change of the order of the chemical reaction taking place \cite{marciano1982curing}. For the case of DBTDL, early studies of Marciano et. al \cite{marciano1982curing} have shown that a minimum concentration of 2.6 mol.m$^{-3}$ was appropriate. On the other hand, the maximum value is usually defined as a function of the processing method. For example, for the case of spray polyurethane foams (SPF), moderate to high concentrations are necessary to achieve a fast surface adhesion of the foam. On the other hand, for the case of thermosets applied in the polymer composite industry, the processing window is usually deduced from chemoviscosity profiles \cite{yuksel2021material}. For example, if infusion \cite{poodts2013fe} is used as the manufacturing process, it might be argued that having an apparent viscosity below 1000 mPa.s is a key prerequisite for the successful application of a thermosetting polymer\cite{mazumdar2001composites}. For this reason, we have reported the chemoviscosity profile of the proposed formulation (see section 4).\\
	\\The effect of catalyst type and concentration was analyzed from the exothermic thermal transitions identified by performing dynamic thermal scans using DSC analysis. These events were quantified by indicating the position of the thermal event (exothermic peak temperature) and the total enthalpy associated to it ($\Delta$H$_{T}$). The results of these analyses are reported in Table 2. For the case of the TsPU$_{1}$, an exothermic event centered at 80\textdegree\ C with an enthalpy of 3.32 J.g$_{-1}$ was identified. The position as well as the enthalpic value clearly indicated that the SOYPOL had a low reactivity, an aspect which can be explained by the fact that the hydroxides present in the polyol are secondary and that the oxirane rings also contribute to a strong steric hindrance effect \cite{ionescu2005chemistry}. If the $\Delta$H$_{T}$ is expressed as a function of isocyanate equivalents (see Table 2), we can also notice that the value is well below the heat of reaction of typical isocyanate-hydroxyl reactions reported in literature \cite{macosko1989rim,zhao2015}. This result was expected and it can be explained by the absence of a catalyst in the formulation as well as the low reactivity of the SOYPOL. FTIR analysis (Fig. 1) of the TsPU$_{1}$ revealed the presence of urethane bonds, hydroxides and free isocyanates, supporting the hypothesis that those reactions had occurred, but to a lesser extent.\\
	\\The effect of increasing amounts of DMCHA on the cure enthalpy of the formulation is also reported in Table 2. The thermal transition shifted to higher temperatures \cite{lipshitz1977kinetics} and the $\Delta$H$_{T}$ increased by +174\% (for a 0.9 wt.\% of DMCHA), indicating that the catalyst was effective in increasing the catalytic activity of the formulation. This fact clearly suggests that the formulation was capable of producing reactions which might not be only associated to urethane linkages. For temperatures above 100\textdegree\ C, a wide variety of reactions between isocyanate groups are feasible \cite{ionescu2005chemistry}. This hypothesis was corroborated by the FTIR analysis (Fig. 1), which corroborated the presence of isocyanurate, urea and other chemical bonds typically found in such formulations.\\
	\\The effect of incorporating DBTDL in the formulation is reported in Table 2 (TsPU$_{5}$). At only 0.2 wt.\%, the thermal transition shifted to much lower temperatures and the $\Delta$H$_{T}$ also increased to similar levels of the TsPU$_{4}$ formulation. This fact clearly indicated that this catalyst was much more effective at lower concentrations, a result which was expected \cite{Randall2002ThePB}.\\
	\\The effect of incorporating glycerin in the formulation catalyzed with DBTDL is reported also in Table 2 (TsPU$_{6}$, TsPU$_{7}$ and TsPU$_{8}$). A substantial increase of the $\Delta$H$_{T}$ was measured, reaching a value of 35.7 J.g$^{-1}$. Taking into account that this formulation had the highest amounts of hydroxide equivalents, it was logical to obtain those results. In addition, the $\Delta$H$_{TNCO}$ (enthalpy normalized with respect to isocyanate equivalents) was also the highest, indicating that the role of urethane linkages was predominant for the formation of a crosslinked network.\\
	\\Finally, the effect of isocyanate index (NCO$_{index}$) is also reported in Table 2. It is important to highlight that the index effect was based on the TsPU$_{7}$ formulation. By reducing the index to 0.31, we found a 62\% decrease of the $\Delta$H$_{T}$, clearly indicating that the formulation had a deficit of isocyanate equivalents. On the other hand, an increase of the NCO$_{index}$ to 1.37 also caused a 13.4\% decrease of the $\Delta$H$_{T}$, indicating an excess of isocyanate equivalents. Taking into account these results, it is logical to deduce that the proposed TsPU$_{7}$ formulation had an adequate stoichiometric relation, giving rise to a balance of hydroxides as well as isocyanate equivalents. This is further confirmed by the $\Delta$H$_{TNCO}$, which was the highest of all the tested formulations.\\
	\\Lastly, it is important to comment on the specific values of $\Delta$H$_{TNCO}$ measured for all the formulations. In scientific literature \cite{castro1982studies,lipshitz1977kinetics}, and, in particular, when model systems are under analysis, the total reaction enthalpy is normalized with respect to the isocyanate or hydroxides equivalents. The main purpose of this normalization is to compare the enthalpies for the formation of a urethane bond. For example, several studies have indicated that, for model systems, the urethane bond formation should be 83.6 .103 J.mol$_{-1}$\cite{macosko1989rim,zhao2015}. Then, by analyzing the results reported in Table 2, it might be argued that the values reported are far below what it has been stablished by previous works. However, it is important to highlight that not only urethane bonds are being formed in the TsPU$_{7}$ formulation. As already discussed above and supported by FTIR analysis (Fig. 1), other chemical bonds were present, which might be associated to lower formation energies. Another important aspect of this formulation has to do with the role of the rubber to glass transition temperature (Tg). As it will be explained in detail in the following section, this formulation had a cure kinetics which was substantially hindered by diffusion effects. This meant that it was logical to have a low $\Delta$H$_{TNCO}$ during cure, because such systems tend to have very long post-cure cycles. This hypothesis was also supported by the results presented in section 3.3.\\

\subsection{The role of polyol crystallization on cure kinetics}
	Due to the fact that the SOYPOL used in this work was synthesized from an epoxidation reaction, it is possible that the unreacted oxirane groups present in the molecular structure of the SOYPOL might lead to macromolecular crystallization \cite{petrovic2007network}. This phenomena has previously been identified by other studies \cite{petrovic2007network,lin2008kinetic}, however, all the melting transitions were found well below -10\textdegree\ C, indicating that this phenomena was relevant only for very low temperatures. It is important to highlight that the authors have not found kinetic studies which quantify the crystallization nor melting of such polyols. As far as our studies are concerned, the crystallization of the SOYPOL was visually identified after samples were stored at temperatures ranging from -18\textdegree\ C to approximately  4\textdegree\ C . The formation of crystals of the SOYPOL as well as the SOYPOL formulated with glycerin are depicted in Fig. 2.\\
	\\At this point, it is important to emphasize the relevance of understanding the concept that crystallization can have a deleterious impact on the final properties of a TsPU. For example, if a polyol that has been partially crystallized is used in the formulation, this means that, under this condition, the polyol will have a lower OH\#. This is because the crystals behave as a second solid phase with a very low reactivity towards the isocyanate precursor. The consequence is that the polyurethane formulation will have a higher NCO$_{index}$, because less hydroxides groups will be available to react with the isocyanate precursor. In addition, due to the fact that the polyurethane cure generates heat, the crystals will melt, giving rise to localized zones with incomplete cure. Such phenomena will certainly affect adversely the thermomechanical properties of the resultant TsPU.\\
	\\To further understand this phenomena, a series of experiments were performed studying the effect of temperature on crystallization. The polyols reported in Table 3 were stored at temperatures within the range of -18 \textdegree\ C to 4\textdegree\ C for periods of time ranging from 15 days to 180 days. Subsequently, samples of those conditioned polyols were analyzed with DSC using a standard dynamic thermal scan so as to identify melting endotherms. For example, for the case of the SOYPOL stored at 4\textdegree\ C for 180 days, an endotherm centered at 44\textdegree\ C and with a total endothermic enthalpy of 4.74 J.g$_{-1}$ was measured. The presence of a melting endotherm centered at temperatures well above ambient temperature clearly suggested that polyol crystallization could have a strong effect on cure kinetics.\\ 

\subsection{Dynamic Mechanical Analysis (DMA)}
	The elastic bending modulus (E’) as well as the damping factor (Tan $\delta$) as a function of temperature for the TsPU$_{7}$ formulation are depicted in Fig. 3. The sample TsPU$_{7}$-1 was obtained from a plate cured at ambient temperature. After this thermal cycle was conducted, the same sample was subjected to a second thermal cycle, denominated as TsPU$_{7}$-2. For the case of the TsPU$_{7}$-1, the storage bending modulus (E’) had a substantial decrease as a function of temperature, starting at 4.66.109 Pa at ambient temperature and going down to 3.58.108 Pa at 180\textdegree\ C. In addition, the damping factor (Tan $\delta$) presented a thermal transition centered at approximately 100\textdegree\ C. However, this transition was not clearly defined, indicating that the sample was curing when the analysis was being conducted. On the other hand, for the case of the TsPU$_{7}$-2, a clearly defined thermal transition (Tg) was found at approximately 220\textdegree\ C, indicating that the material attained a higher conversion. In addition, the residual storage elastic modulus was 6.7.107 Pa, indicating that only a 1.3\% of the initial E’ was retained at 240\textdegree\ C.\\
	\\It is important to highlight that, as already mentioned in the previous section, vitrification inhibited substantially the final conversion of the TsPU. When a sample is obtained from an unheated mould, it is logical to expect that a post-cure cycle will be necessary to achieve improved thermo-mechanical properties. We chose to conduct these experiments intentionally so as to highlight the impact of vitrification on the thermomechanical properties of the TsPU$_{7}$.\\

\subsection{Quasi-static flexural mechanical tests}
	The results of the flexural mechanical tests performed on samples of the TsPU$_{7}$ post-cured at 70\textdegree\ C (TsPU$_{7}$ – 70\textdegree\ C) and 110\textdegree\ C (TsPU$_{7}$ – 110\textdegree\ C) are reported in Table 4. The flexural strengths attained under both conditions were similar, with a maximum value of 99.4 MPa for the case of the sample cured at a higher temperature. The flexural modulus had a strong change as a function of post-cure cycle, increasing by up to 24\% to 2.14 GPa for the case of the TsPU$_{7}$ – 110\textdegree\ C. In addition, the standard deviation of the flexural modulus decreased substantially as a function of higher post-cure cycle temperatures. This was a clear indication that supported the hypothesis that vitrification was playing a key role in the post-cure of this thermosetting formulation. For this reason, higher temperatures were needed in order to attain full cure and, subsequently, to homogenize the mechanical properties of the material under analysis. Finally, the flexural strain to failure was also highly dependent on post-cure cycle, decreasing by up to 30\% to 5.69\% for the case of high temperature post-cure cycle. This result also reflected the fact that subsequent crosslinking took place under the post-cure cycle. Such effect was expected due to the fact that the initial crosslinking of the low OH\# polyol (ESO) caused vitrification (see section 4.2), decreasing substantially the rate of reaction of the OH groups present in the glycerin polyol.\\
	
\section{Chemorheological master model of the TsPU$_{7}$ baseline formulation}
\subsection{Cure kinetics of the TsPU$_{7}$ formulation}
	The heat flow as a function of time for isothermal experiments during the cure of the TsPU$_{7}$ are depicted in Fig. 4. The isothermal experiments were performed from 30\textdegree\ C all the way up to 70\textdegree\ C. First, it is important to emphasize that the time was synchronized with the start of the DSC isothermal experiment. The total time should also include the sampling preparation time, which is reported in the experimental section (2.1). This preparation time sets the limit of the maximum isothermal temperature, which was 70\textdegree\ C. For higher temperatures, the rise of the speed of the reaction would increase the exothermic heat flow to such an extent that its measurement would not be possible.\\
	\\The total reaction enthalpy ($\Delta H_{ISO}$) associated to each isothermal experiment is reported in table 5. Taking into account the $\Delta H_{T}$ reported in table 2, the maximum conversion ($\alpha_{max}$) of each isothermal experiment was also calculated and reported in table 5. For example, at 30\textdegree\ C, a maximum conversion of 42\% was achieved during cure. Clearly, this was an indication that vitrification took place and that a proper cure cycle should certainly include a post-cure cycle. On the other hand, the highest isothermal temperatures indicated a full conversion, indicating that full cure was attained within this temperature range.\\
	\\The evolution of the maximum conversion as a function of cure temperature was found to follow a Boltzmann behavior according to the following equation\ :\\
	
	\begin{equation}
\label{eq:sedov}
\alpha_{max}=\frac{A}{1+\exp(\beta(T-T_{0.5})})+B
\end{equation}\\

    Where A, $\beta$ were fitting parameters which affected the slope of the conversion curve and T$_{0.5}$ was the absolute temperature (K) at which half conversion was achieved. The fitted parameters can be consulted in table 6.\\
    \\To better understand which phenomenological model was appropriate to predict the cure kinetics of the TsPU$_{7}$, the experimental data presented in Fig. 5 was expressed as a function of conversion rate using equation eq. 2, where $\Delta H_{ISO}$ was the total enthalpic contribution associated to each isothermal experiment and $\Delta H_{T}$ represented the total enthalpy of the TsPU$_{7}$ formulation. The experiments expressed using conversion are reported in Fig. 6. The shape of the curves presented in Fig. 5 were used to infer that an autocalytic phenomenological model should be implemented for the TsPU$_{7}$. In addition, taking into account that the maximum conversion increased as a function of temperature, it was also necessary to include vitrification in the model. Hence, the following equation was proposed,\\
    
    \begin{equation}
\label{eq:sedov}
\frac{d\alpha}{dt}=k\alpha^m(\alpha_{max}-\alpha)^n
\end{equation}\\

    Where n and m represented the reaction orders and k is the frequency factor that includes an Arrhenius type temperature dependency and is determined as follows,
    
    \begin{equation}
\label{eq:sedov}
k=k_0\exp^{(\frac{E_a}{RT})}
\end{equation}\\

Where k$_{0}$ is the pre-exponential factor, E$_{a}$ is the activation energy, R is the gas constant and T is the absolute temperature. The activation energy (E$_{a}$) was calculated from the slope of the relationship between inverse temperature (1/T) and natural logarithm of the frequency factor previously fitted in the rate of conversion (d$\alpha$/dt) for each temperature. All values were reported in table 7 and table 8.\\

\subsection{Evolution of rubbery to glass transition temperature (Tg) as a function of conversion}
    The evolution of Tg as a function of conversion is a key aspect to understand the role of vitrification on the cure of a thermosetting polymer \cite{teil2004ttt}. From a processing point of view, the occurrence of vitrification represents a drawback, because it limits the maximum conversion during cure, affecting the final mechanical properties of the polymer under analysis. A glassy state during cure will certainly indicate that a post-cure cycle should be implemented, so as to attain the maximum Tg of the thermosetting polymer. However, vitrification can also be used to induce a cure inhibition effect, due to the fact that the glassy state inhibits the formation of covalent bonds within the reactive mixture. Further details about the role of vitrification on poly(urethane-isocyanurate) thermosets can be consulted in a previous study of our group \cite{chiacchiarelli2013kinetic}.\\
    \\To understand how the Tg evolved as a function of conversion, dynamic thermal experiments where performed on samples previously cured at isothermal conditions. The evolution of Tg as a function of conversion can be depicted in Fig. 7. The Di-Benedetto equation was used to model the experimental results using the following equation,\\
    
    \begin{equation}
\label{eq:sedov}
\frac{Tg-Tg_0}{Tg_0}=\frac{(\frac{E_\infty}{E_0}-\frac{C_\infty}{C_0})\alpha}{1-(1-\frac{C_\infty}{C_0})\alpha}
\end{equation}\\

    where Tg$_0$ indicates transition temperature of the initial monomers in the system, $\alpha$ extent of reaction (conversion), E$_{\alpha}$/E$_{0}$ lattice energy of cross-linked and partial cross-linked polymer ratio and C$_{\alpha}$/C$_{0}$ the segment mobility ratio.\\
    \\The evolution of Tg as a function of conversion is a key measurement so as to understand the role of vitrification on the cure kinetics of the thermosetting polymer under analysis. To better understand this, we have incorporated in Fig. 7 the results of other thermosetting polymers, particularly poly(urethane-isocyanurate) [10] and anhydride cured epoxy systems \cite{belnoue2016novel}. For a fixed conversion, i.e. 40\%, we can deduce that the Tg of polyurethane systems was much higher with respect to epoxy. This tendency was also found for the case of low conversions, but, for conversions above 80\%, the Tg’s tended to converge to similar values. These observations clearly reflect how vitrification influences the cure of polyurethane and epoxy systems. Whenever a thermosetting polymer presents a Tg versus conversion curve shifted upwards (Fig. 7), then, it is expected that vitrification will certainly inhibit cure. This is because the gelation process is much slower in comparison to the formation of molecular chains which can undergo vitrification at curing temperatures. Hence, it is also expected that the final cure of the polymer will also take much more time, because the chemical reactions are mostly taking place under a glassy state. However, this effect can also be used to inhibit cure, an aspect which is fundamental for the development of prepregs \cite{centea2015review,tuncay2018fast}.\\

   \subsection{Chemorheological behavior of the TsPU$_{7}$}  
   The apparent viscosity as a function of time for isothermal experiments of the TSPU$_{7}$ formulations are depicted in Fig. 8. For an isothermal experiment at 40\textdegree\ C, the initial apparent viscosity ranged at 5.10$^{2}$ cp and started to increase significantly after 25 min., reaching 3.10$^{3}$ mPa.s at approximately 45 min. In contrast, the isothermal experiment at 60\textdegree\ C yielded an apparent viscosity of 3.10$^{3}$ mPa.s at only 12.5 min. To better understand the evolution of apparent viscosity as a function of conversion, the results depicted in Fig. 9 were modeled using the following equation, proposed originally by Kim and Macosko \cite{macosko1989rim},\\
   
    \begin{equation}
\label{eq:sedov}
\eta (T,\alpha)=\eta (T)(\frac{\alpha_g}{\alpha_g-\alpha})^{a+b\alpha}
\end{equation}\\

     where “$\alpha$g” is the theoretical gel point ($\alpha$g = 0.533), “a” and “b” are fitting parameters and the function $\eta$(T) has an Arrhenius dependence as indicated in the following equation,\\
     
    \begin{equation}
\label{eq:sedov}
\eta(T)=\eta_0\exp^{(\frac{E_a}{RT})}
\end{equation}\\

     where $\eta_{0}$ is the pre-exponential factor, and E$_{a,M}$ is the activation energy, R is the gas constant and T is the absolute temperature. The fitting parameters obtained from the experimental results are reported in table 9.\\
     \\A very important deduction from Fig. 9 has to do with the role of vitrification and gelation in the evolution of apparent viscosity. For example, for the case of an isothermal experiment at 40\textdegree\ C, from Fig. 9 it can be understood that at a conversion of approximately 0.25, the increase of apparent viscosity steeped up significantly. Due to the fact that the theoretical gelation point was located at much higher conversions (0.53), it can be deduced that vitrification played the most important role in the increase of apparent viscosity. These results support what has been found in previous sections, that is, vitrification was the main cause of the observed steep increase in apparent viscosity. Then, it is logical that subsequent post-cure cycles should be performed so as to achieve higher chemical conversions.\\
     \\Another important aspect which needs to be discussed has to do with mass transfer effects. On the one hand, cure reactions are exothermic, so, it is expected that heat flow will have to be dissipated according to the specific geometry of the experiment being conducted. If a high area per unit volume experiment is performed, a lower adiabatic medium shall be present and the material would maintain its original temperature to a much better extent. On the other hand, if isothermal experiments are performed, heat exchange will inevitable affect the reaction kinetics of the polymer system being analyzed \cite{pascault2002thermosetting}. These effects have been extensively studied in literature, but it is important to emphasize the fact that it will inevitably affect any rheological experiment being implemented. For example, in this work, we can deduct from the in-situ temperature measurements (Fig. 8) that mass effects contributed to an increase in the time to reach the specific isothermal temperature that was priory stablished. It might be argued that a different sample mass should be implemented to avoid this, but, the experiments were performed according to standard recommendations regarding the measurement of apparent viscosity of thermosetting polymers. It is extremely important to report the results presented in Fig. 8 because those are important for the application of this thermosetting in polymer composites. From a strictly modeling point of view, it is clear that a parallel plate rheometer [4] will provide a much better set of experimental data, but we also think it is very important to conduct and report rotational experiments because these are widely employed in industry.\\
     
\section{Conclusions}
     A thermosetting polyurethane was synthesized by combining an aromatic isocyanate and a polyol obtained from soybean oil using ESO crosslinked with glycerin. Cure kinetics analysis with DSC revealed that DBTDL was the most effective catalyst for the proposed formulation, with a total reaction enthalpy of 35.7 J.g$^{-1}$. Aminic catalysts improved reaction kinetics, but to a lesser extent. DSC analysis were able to corroborate that a proper NCO$_{index}$ was selected, maximizing the reaction enthalpy for the proposed formulation. A cure kinetics model based on autocatalytic heat flow where vitrification was preponderant in the evolution of conversion was obtained. FTIR analysis of cured samples corroborated the formation of both urethane as well as isocyanurate and urea bonds. The evolution of Tg’s as a function of temperature corroborated that vitrification slowed down cure kinetics, particularly at isothermal experiments performed at lower temperatures. ESO crystallization was induced and quantified with DSC, finding a melting endotherm well above room temperature. This clearly indicated that ESO crystallization might have a strong effect on cure kinetics. DMA analysis of un-cured and in-situ post-cured samples revealed that a Tg centered at 220\textdegree\ C was attained. Quasi-static flexural mechanical tests proved that post-cure cycles had a strong effect on flexural modulus, strain to failure and strength. A maximum flexural modulus of 2.14GPa was obtained for the highest temperature post-cure (110\textdegree\ C) and a maximum strain to failure of 8.14 \% was obtained for the lowest post-cure temperature (70\textdegree\ C).\\
     \\The previous results emphasize the feasibility of obtaining thermosetting polymers from a soybean oil based polyol. These results will serve to further extend the use of biobased polymers in the polymer composites industry. Ongoing work in this area will focus on the effect of increasing amounts of crosslinking on the thermomechanical properties of the resultant thermosetting polymers, focusing on the role of cure and post-cure cycles.\\   

\textbf{Acknowledgements}\\
\\The author would like to thanks colleagues which indirectly contributed to this work, Matias Ferreyra (Huntsman), Diego Judas (Alkanos) and Hernan Bertolotto (Evonik). The work was supported by the ‘PICT 2015 N0475’ erogated by the ANPCYT (Argentina) as well as the “PIP-2015-N0425”, erogated by CONICET.\\

\bibliographystyle{unsrt}
\bibliography{papercurekinetics}

\begin{table}[h]
\caption{Polyurethane formulations used in this work.} 
\centering 
\begin{tabular}{c c c c c c c} 
\hline 
\\[-1ex]
Formulation & Polyol (pbw) & Isocyanate (pbw) & Glycerin (pbw) & DMCHA (pbw) & DBTDL (pbw) & Isocyanate NCO$_{index}$\\ [0.5ex] 
\hline 
\\[-1ex]
TsPU$_{1}$ & 100 & 36.6 & - & - & - & 1.11\\ [1ex] 
TsPU$_{2}$ & 100 & 36.6 & - & 0.3 & - & 1.11\\
TsPU$_{3}$ & 100 & 36.6 & - & 0.6 & - & 1.11\\
TsPU$_{4}$ & 100 & 36.6 & - & 0.9 & - & 1.11\\
TsPU$_{5}$ & 100 & 36.6 & - & - & 0.2 & 1.11\\
TsPU$_{6}$ & 100 & 36.6 & 20 & - & 0.2 & 0.31\\
TsPU$_{7}$ & 100 & 126 & 20 & - & 0.2 & 1.05\\
TsPU$_{8}$ & 100 & 164 & 20 & - & 0.2 & 1.37\\ [1ex] 
\hline 
\end{tabular}
\label{table:Table1} 
\end{table}

\begin{table}[h]
\caption{Results of the dynamic scans for the formulations used in this work.} 
\centering 
\begin{tabular}{c c c c} 
\hline 
\\[-1ex]
Formulation & $\Delta H_{T}$(J.g$^{-1}$) & $\Delta H_{Tnco}$(J.eq$^{-1}$) & Peak Temperature (\textdegree C)\\ [1ex] 
\hline 
\\[-1ex]
TsPU$_{1}$ & 3.32.10$^{0}$ $\pm$0.6 & 1.68.10$^{3}$ & 7.98.10$^{1}$ $\pm$1.76\\[1ex] 
TsPU$_{2}$ & 4.27.10$^{0}$ & 2.16.10$^{3}$ & 1.38.10$^{2}$\\[1ex]
TsPU$_{3}$ & 4.65.10$^{0}$ & 2.35.10$^{3}$ & 1.38.10$^{2}$\\[1ex]
TsPU$_{4}$ & 9.09.10$^{0}$ & 4.60.10$^{3}$ & 1.38.10$^{2}$\\[1ex]
TsPU$_{5}$ & 9.06.10$^{0}$ $\pm$0.95 & 4.60.10$^{3}$ & 3.32.10$^{1}$ $\pm$3.48\\[1ex]
TsPU$_{6}$ & 1.36.10$^{1}$ $\pm$5.2 & 7.89.10$^{3}$ & 3.55.10$^{1}$ $\pm$1.4\\[1ex]
TsPU$_{7}$ & 3.57.10$^{1}$ $\pm$3.45 & 9.45.10$^{3}$ & 3.76.10$^{1}$ $\pm$2.68\\[1ex]
TsPU$_{8}$ & 3.09.10$^{1}$ & 7.25.10$^{3}$ & 3.85.10$^{1}$\\ [1ex] 
\hline 
\end{tabular}
\label{table:Table2} 
\end{table}

\begin{table}[h]
\caption{Endothermic enthalpy as a function of exposure time/temperature for SOYPOL.} 
\centering 
\begin{tabular}{c c c c c} 
\hline 
\\[-1ex]
Formulation & Storage time (days) & Temperature (\textdegree C) & Endothermic enthalpy (J.eq$^{-1}$) & Tp (\textdegree C)\\ [1ex] 
\hline 
\\[-1ex]
SOYPOL & 180 & 4 & 4.47 & 44\\[1ex] 
SOYPOL & 90 & -18 & 0.32 & 36\\[1ex] 
\hline 
\end{tabular}
\label{table:Table3} 
\end{table}

\begin{table}[h]
\caption{Quasi-static flexural properties of the TsPU$_{7}$.} 
\centering 
\begin{tabular}{c c c c c} 
\hline 
\\[-1ex]
Material & Flexural Strength (Mpa) & Flexural Modulus (Gpa) & Flexural Strain to Failure (\%)\\ [1ex] 
\hline 
\\[-1ex]
TsPU$_{7}$ – 70\textdegree C & 90.7 $\pm$18.1 & 1.73 $\pm$0.537 & 8.10 $\pm$1.69\\[1ex] 
TsPU$_{7}$ – 110\textdegree C & 90.7 $\pm$18.1 & 1.73 $\pm$0.537 & 8.10 $\pm$1.69\\[1ex] 
\hline 
\end{tabular}
\label{table:Table4} 
\end{table}

\begin{table}[h]
\caption{Cure enthalpy and maximum conversion for isothermal experiments of TsPU$_{7}$.} 
\centering 
\begin{tabular}{c c c} 
\hline 
\\[-1ex]
Temperature (\textdegree C) & $\Delta H_{iso}$(J.g$^{-1}$) & $\alpha_{max}$\\ [1ex] 
\hline 
\\[-1ex]
30 & 15 & 0.42\\[1ex] 
40 & 22.1 $\pm$1.93 & 0.62 $\pm$0.05\\[1ex]
50 & 26.4 $\pm$2.7 & 0.74 $\pm$0.07\\[1ex]
60 & 30.3 $\pm$2.15 & 0.85 $\pm$0.06\\[1ex]
70 & 35.7 $\pm$3.31 & 1 $\pm$0.09\\[1ex]
\hline 
\end{tabular}
\label{table:Table5} 
\end{table}

\begin{table}[h]
\caption{Fitting parameters for the maximum conversion as a function of temperature for TsPU$_{7}$.} 
\centering 
\begin{tabular}{c c} 
\hline 
\\[-1ex]
Parameter & Neat\\ [1ex] 
\hline 
\\[-1ex]
A & -1.765\\[1ex] 
B & 1.381\\[1ex] 
T$_{0.5}$ & 33.99 [\textdegree C]\\[1ex]
$\beta$ & 0.0349 [1/\textdegree C]\\[1ex]
Corr. coeff. & 0.9814\\[1ex]
\hline 
\end{tabular}
\label{table:Table6} 
\end{table}

\begin{table}[h]
\caption{Fitting parameters of the cure kinetic model of TsPU$_{7}$.} 
\centering 
\begin{tabular}{c c c c c c} 
\hline 
\\[-1ex]
Temperature (\textdegree C) & k (1/s) & n & m & n+m & Corr. coeff.\\ [1ex] 
\hline 
\\[-1ex]
30 & 0.0122 & & & & \\[1ex] 
40 & 0.0112 & & & & \\[1ex] 
50 & 0.018 & \raisebox{1ex}{1.70} & \raisebox{1ex}{0.47} & \raisebox{1ex}{2.17} & \raisebox{1ex}{0.9847}\\[1ex]
60 & 0.0326 & & & & \\[1ex]
70 & 0.0336 & & & & \\[1ex]
\hline 
\end{tabular}
\label{table:Table7} 
\end{table}

\begin{table}[h]
\caption{Fitting parameters obtained by the Arrhenius equation.} 
\centering 
\begin{tabular}{c c c c} 
\hline 
\\[-1ex]
k  & T (\textdegree K) & E$_{a}$ (kJ.mol$^{-1}$) & R$_{a}$ (J.molK$^{-1}$)\\ [1ex] 
\hline
\\[-1ex]
0.012 & 303.15 & & \\[1ex] 
0.011 & 313.15 & & \\[1ex] 
0.018 & 323.15 & \raisebox{1ex}{27.05} & \raisebox{1ex}{8.31}\\[1ex]
0.0326 & 333.15 & & \\[1ex]
0.0336 & 343.15 & & \\[1ex]
\hline 
\end{tabular}
\label{table:Table8} 
\end{table}

\begin{table}[h]
\caption{Fitting parameters obtained by the rheological model of TsPU$_{7}$.} 
\centering 
\begin{tabular}{c c c c c} 
\hline 
\\[-1ex]
$\alpha_{g}$  & $\eta_{0}$ (mPa.s) & E$_{a,M}$ (kJ.mol$^{-1}$) & a & b\\ [1ex] 
\hline
\\[-1ex]
0.528 & 2.538.10$^{-5}$ & 61.43 & 2.00 & -1.58\\[1ex] 
\hline 
\end{tabular}
\label{table:Table9} 
\end{table}

\begin{figure}[h]
        \center{\includegraphics[width=\textwidth]
        {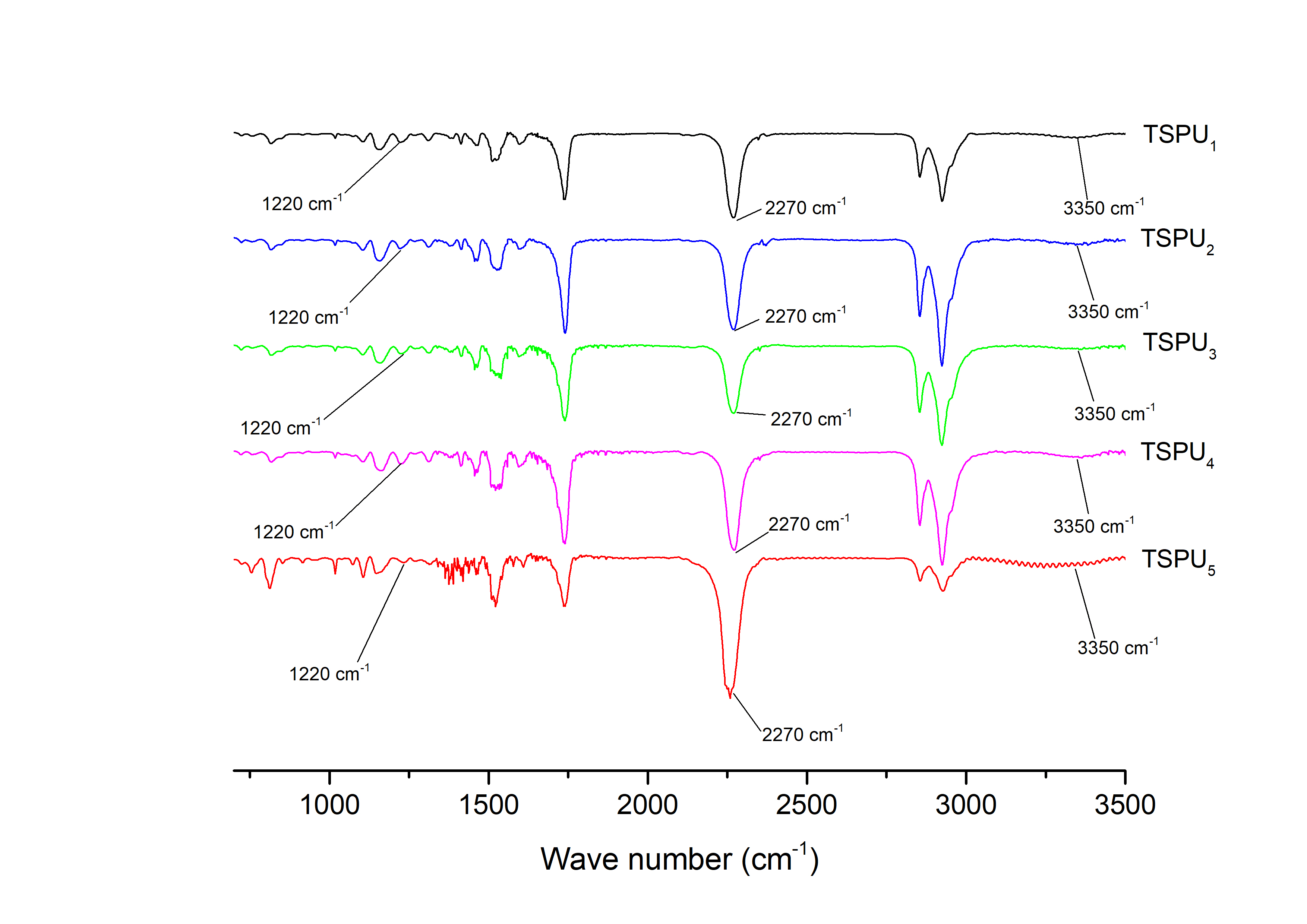}}
        \caption{\label{fig:Fig1} FTIR analysis of the TsPU formulations used in this work.}
      \end{figure}
      
\begin{figure}[h]
        \center{\includegraphics[width=\textwidth]
        {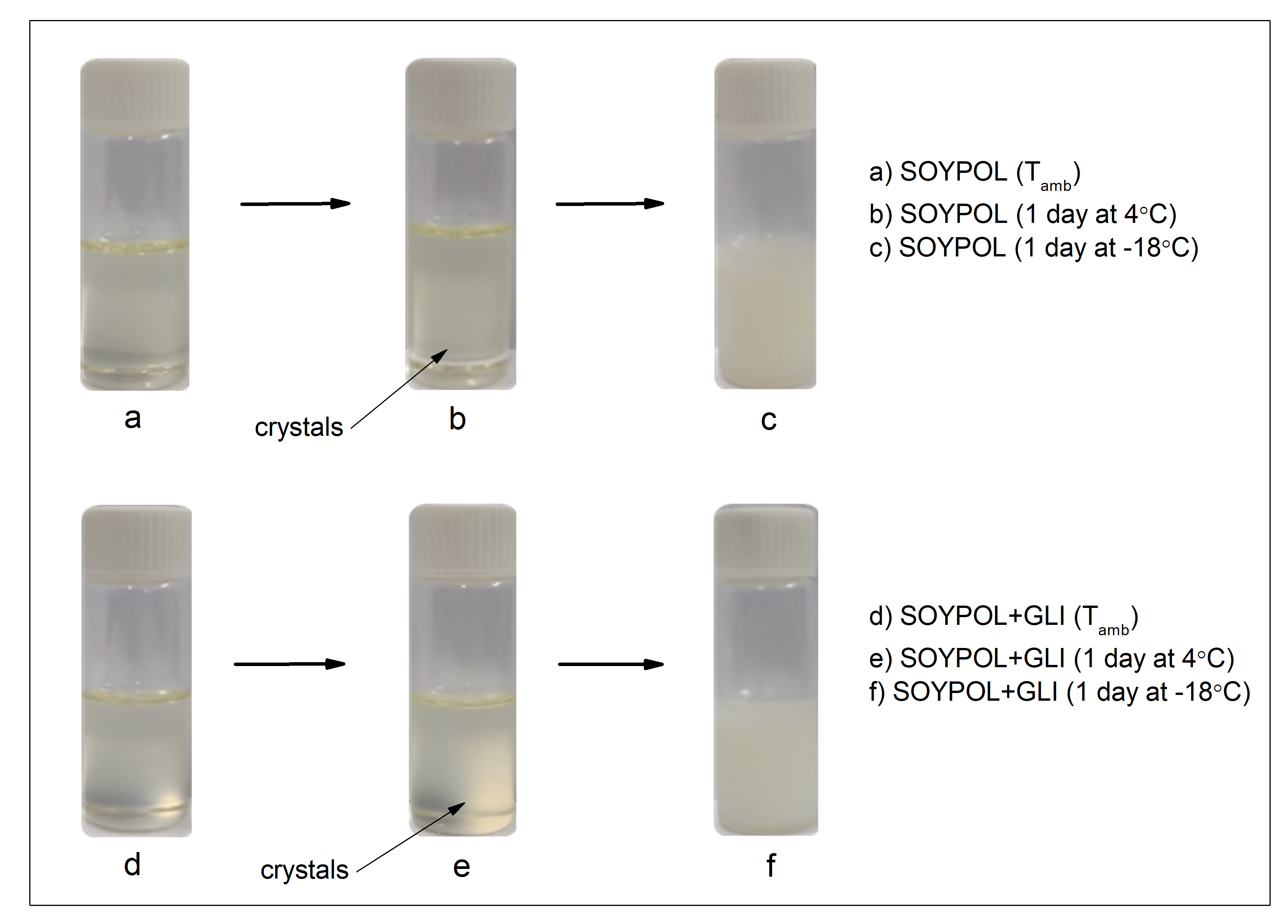}}
        \caption{\label{fig:Fig2} Crystallization of SOYPOL (a, b and c) and SOYPOL + glycerin (d, e and f).}
      \end{figure}
      
\begin{figure}[h]
        \center{\includegraphics[width=\textwidth]
        {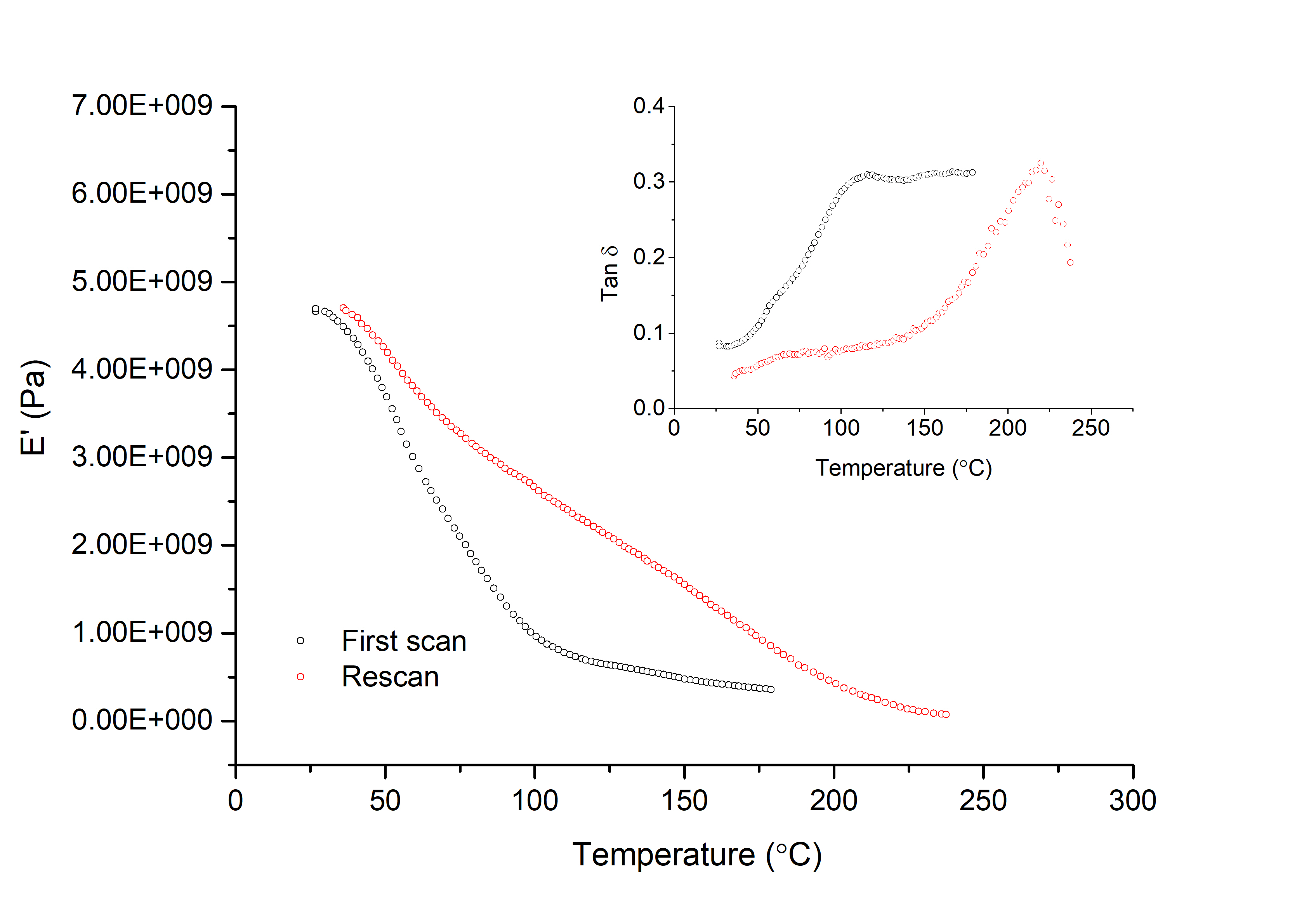}}
        \caption{\label{fig:Fig3} Dynamic mechanical analysis of TsPU$_{7}$. A first scan and re-scan of the same sample were performed.}
      \end{figure}
      
\begin{figure}[h]
        \center{\includegraphics[width=\textwidth]
        {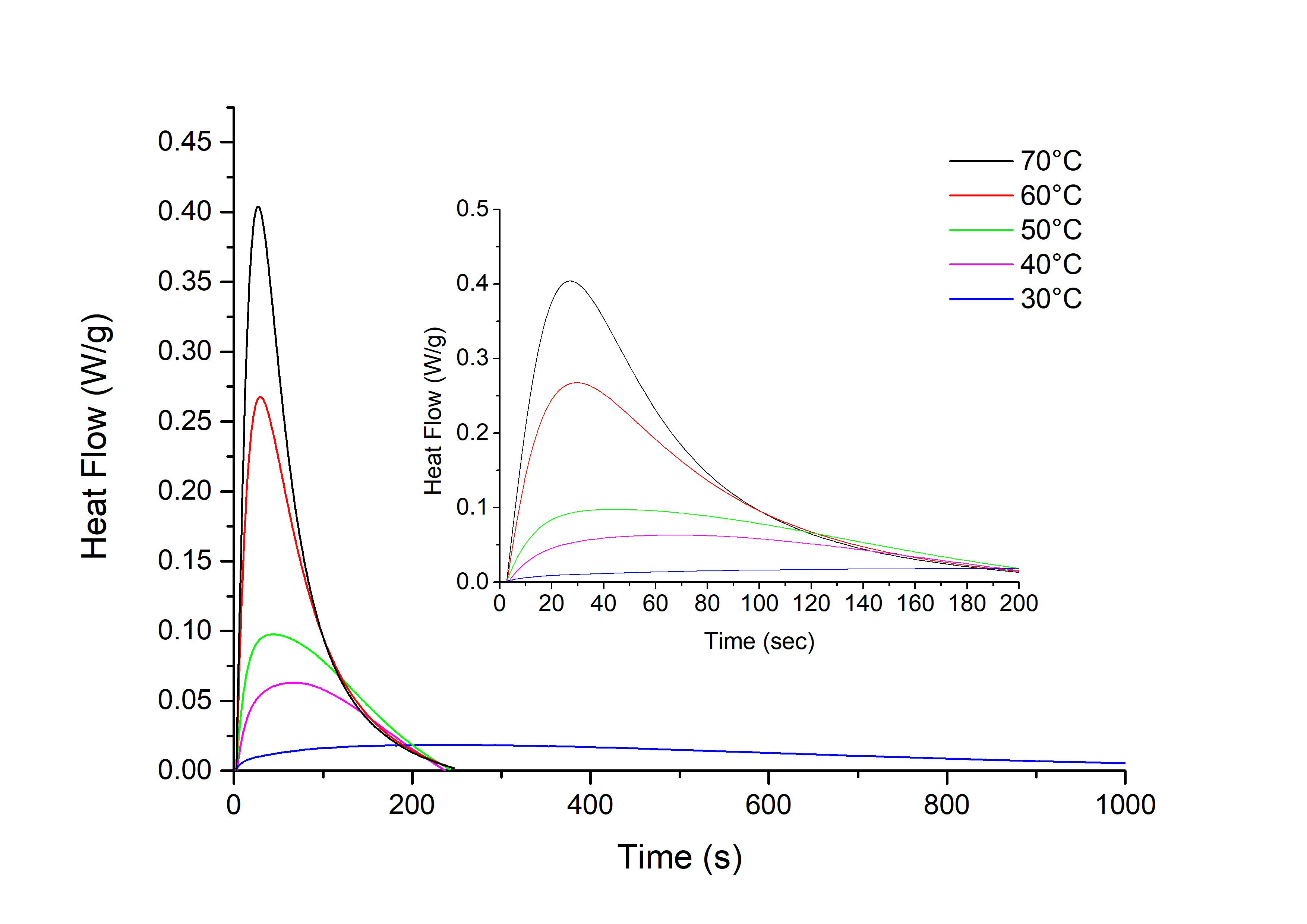}}
        \caption{\label{fig:Fig4} Representative isothermal DSC scans of TsPU$_{7}$.}
      \end{figure}
      
\begin{figure}[h]
        \center{\includegraphics[width=\textwidth]
        {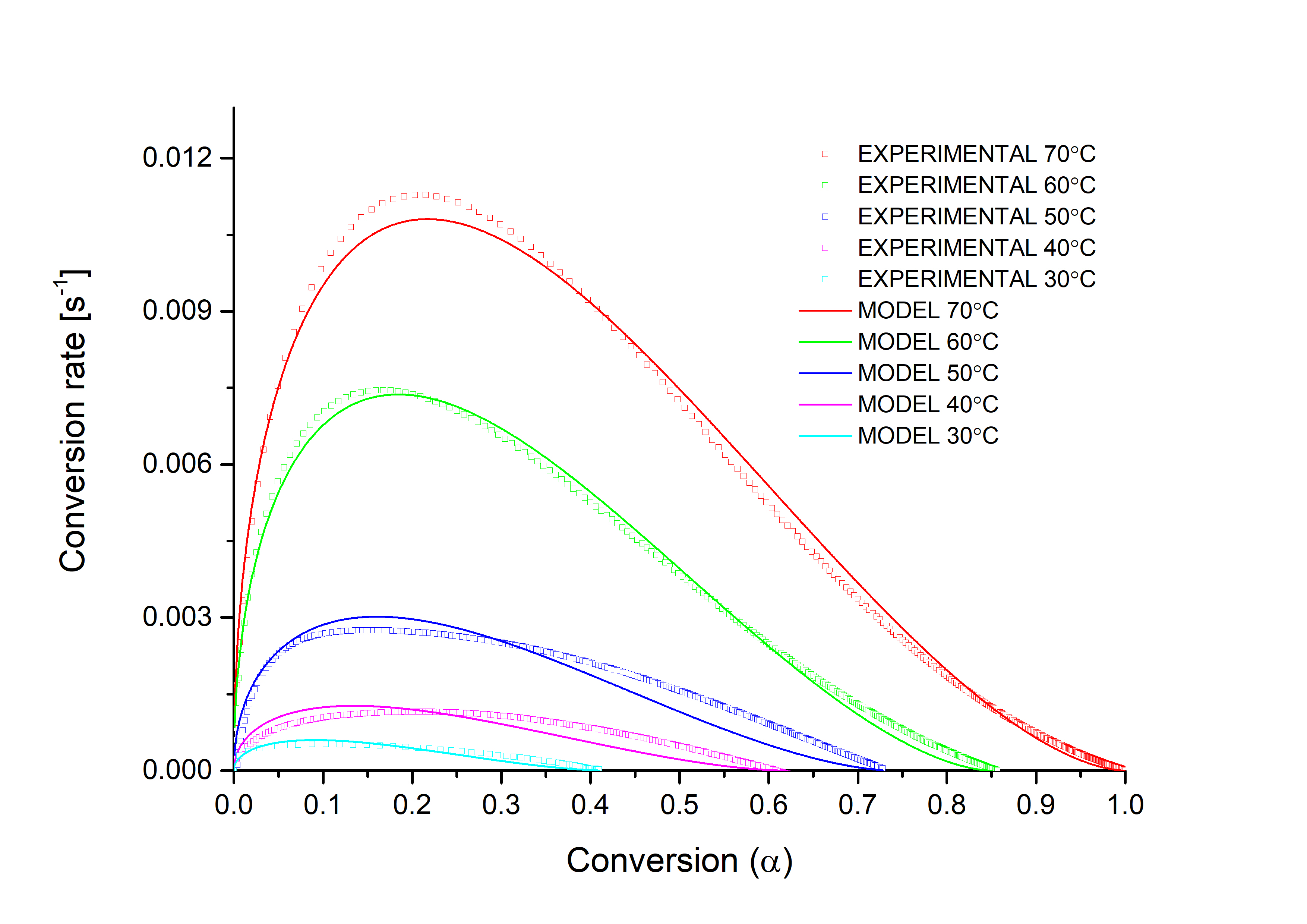}}
        \caption{\label{fig:Fig5} Measured (dashed) and modeled (continuous) conversion rate versus conversion for isothermal conditions at 30\textdegree\ C, 40\textdegree\ C, 50\textdegree\ C, 60\textdegree\ C and 70\textdegree\ C of TsPU$_{7}$.}
      \end{figure}

\begin{figure}[h]
        \center{\includegraphics[width=\textwidth]
        {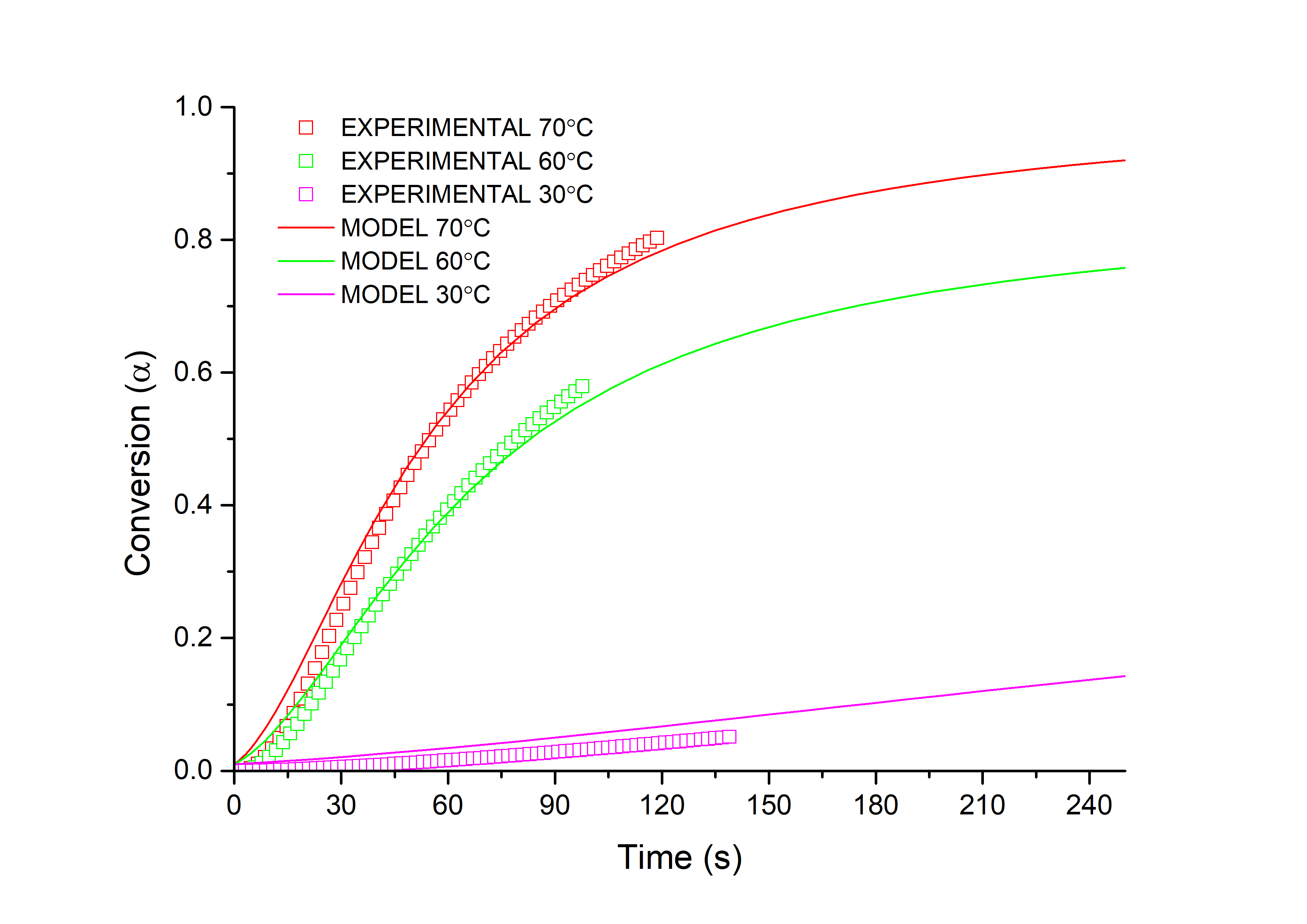}}
        \caption{\label{fig:Fig6} Comparison between experimental conversion against theoretical conversion versus time at 30\textdegree\ C, 60\textdegree\ C and 70\textdegree\ C isothermal conditions of TsPU$_{7}$.}
      \end{figure}
      
\begin{figure}[h]
        \center{\includegraphics[width=\textwidth]
        {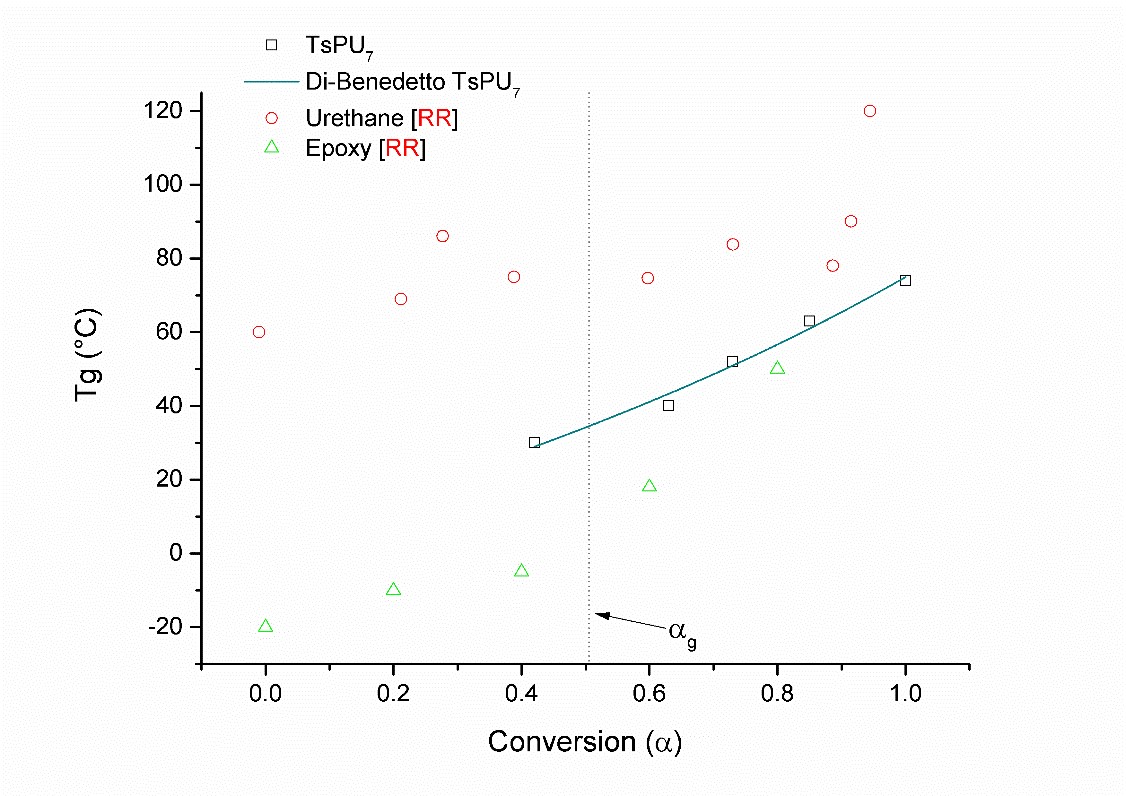}}
        \caption{\label{fig:Fig7} Transition temperature as a function of conversion of TsPU$_{7}$, poly(urethane-isocyanurate) \cite{chiacchiarelli2013kinetic} and anhydride cured epoxy systems \cite{teil2004ttt}}.
      \end{figure}

\begin{figure}[h]
        \center{\includegraphics[width=\textwidth]
        {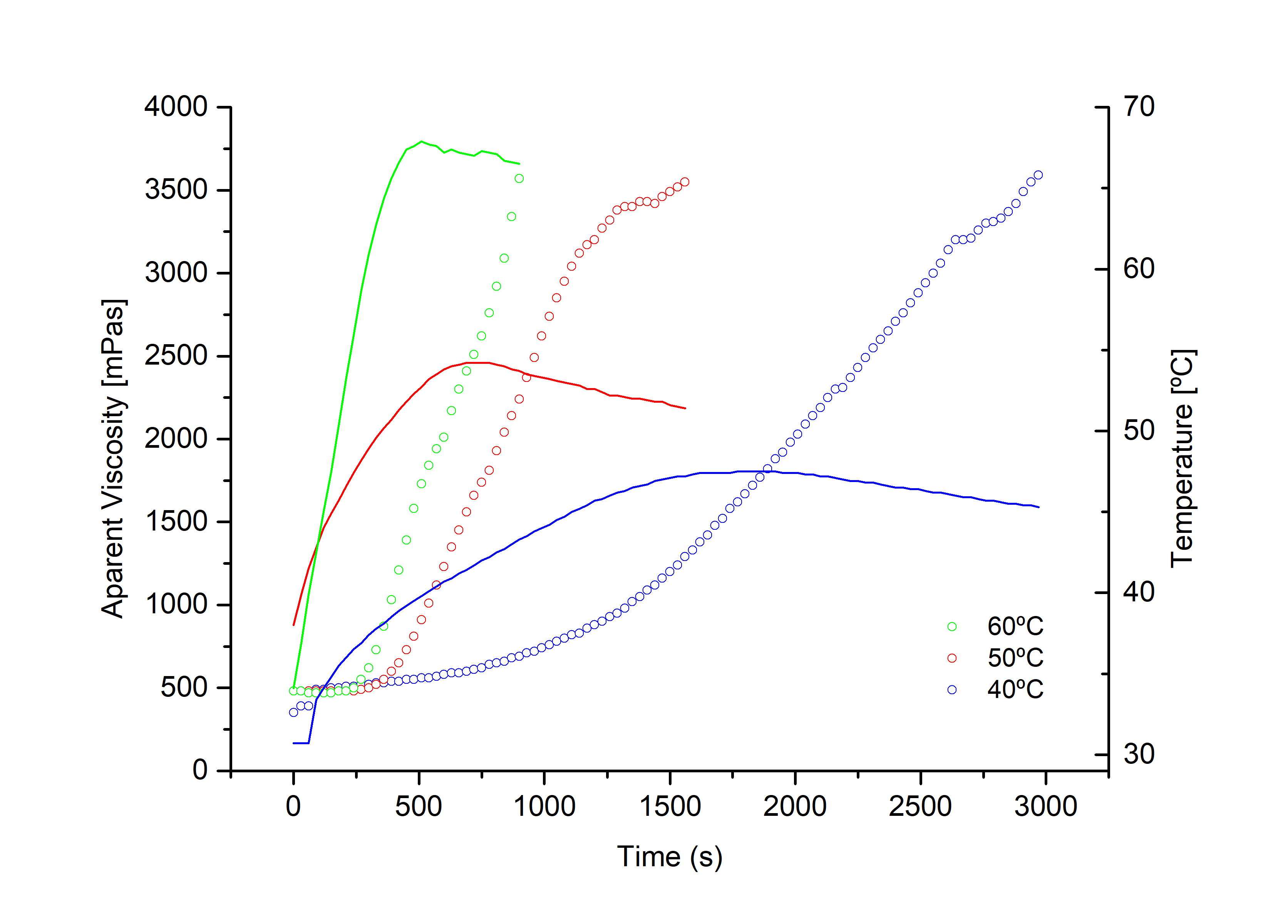}}
        \caption{\label{fig:Fig8} Apparent viscosity as a function of time for 40\textdegree\ C, 50\textdegree\ C and 60\textdegree\ C isothermal cure temperatures of TsPU$_{7}$.}
      \end{figure}
      
\begin{figure}[h]
        \center{\includegraphics[width=\textwidth]
        {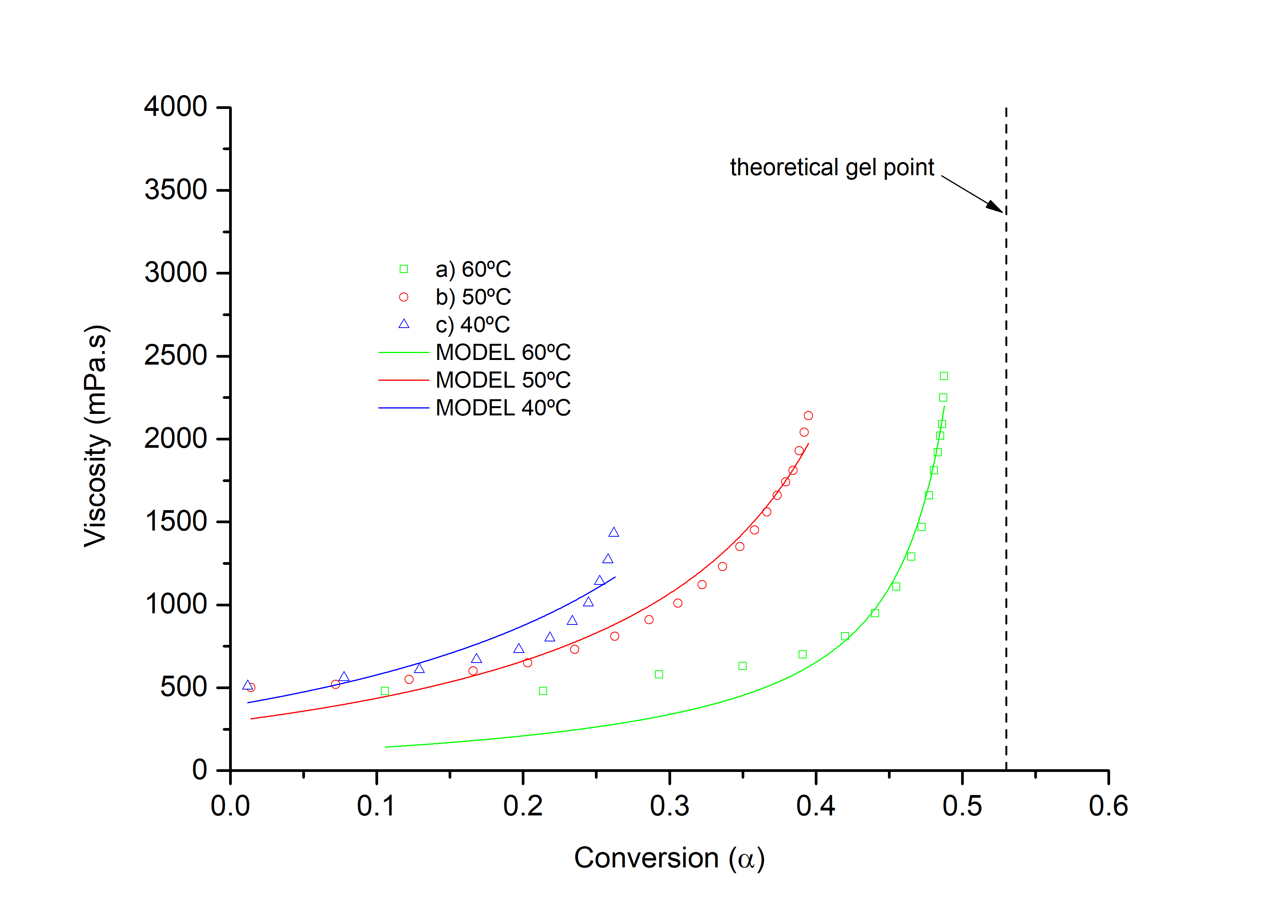}}
        \caption{\label{fig:Fig9} Apparent viscosity as a function of conversion for isothermal temperatures a)60\textdegree\ C, b)50\textdegree\ C and c)40\textdegree\ C of TsPU$_{7}$.}
      \end{figure}

\end{document}